# SPEAKER ATTRIBUTED AUTOMATIC SPEECH RECOGNITION USING SPEECH AWARE LLMS

*Hagai Aronowitz*, Zvi Kons*, Avihu Dekel, George Saon, Ron Hoory*

IBM Research

## ABSTRACT

Speaker-Attributed Automatic Speech Recognition (SAA) enhances traditional ASR systems by incorporating relative speaker identity tags directly into the transcript (e.g., [Speaker 1]:, [Speaker 2]:). In this work, we extend the capabilities of Granite-speech, a state-of-the-art speech-aware Large Language Model (LLM) originally trained for transcription and translation. We demonstrate that it can be effectively adapted for SAA with only minimal architectural changes. Our core contribution is the introduction of speaker cluster identification tags (e.g., [Speaker 1 cluster 42]:) which are jointly trained with SAA to significantly improve accuracy. To address limitations in training data, we propose a data augmentation method that uses artificially concatenated multi-speaker conversations. Our approach is evaluated across multiple benchmarks and shows superior performance compared to conventional pipelines that sequentially perform speaker diarization followed by ASR.

*Index Terms*— SAA, speaker attributed ASR, speaker diarization, speech-aware LLM, Granite Speech

## 1. INTRODUCTION

Automatic Speech Recognition (ASR) has witnessed remarkable advancements in recent years, driven by large-scale pretraining and powerful neural sequence models. Despite these successes, conventional ASR systems are limited to transcribing *what was said*, without identifying *who said it*. For real-world tasks such as meeting transcription, conversational analytics, and dialogue systems, it is imperative to synchronously capture both the spoken content and the speaker identity in order to fully characterize conversational interaction and attribute information correctly.

Speaker-Attributed ASR (SAA) addresses this gap by embedding speaker labels directly into the transcript. Traditional approaches to SAA typically employ multi-stage pipelines, where speaker diarization (SD) performs speaker segmentation and speaker clustering before passing the resulting audio segments to ASR modules [1]. Although these cascaded systems are effective in some scenarios, they often propagate errors across modules and increase system complexity. Another widespread strategy decouples recognition and diarization, aligning word-level timestamps post hoc [2]. However, this is problematic with modern end-to-end ASR architectures, which seldom output precise word timings and miss out on the benefits of joint task optimization.

Unified frameworks designed to perform recognition and speaker attribution simultaneously offer a more robust approach. For example, Shafey et al. [3] addressed the task of transcribing clinical conversations between physicians and patients by training an RNN-T system with extended target symbols indicating speaker roles (*physician* and *patient*).

Recently, speech-aware large language models (LLMs) such as SALMONN [4], SLAM [5], and QWEN-AUDIO [6] have enabled new paradigms for joint speech and speaker modeling. These models combine robust pre-trained speech encoders with the contextual reasoning power of instruction-tuned LLMs. Speech is encoded and projected into the embedding space of a text LLM, where it is concatenated to a text prompt.

However, current speech encoders remain suboptimal for speaker-related tasks, resulting in competitive transcription but limited speaker discrimination. Tasks like speaker verification [4] and speaker counting [7] have been evaluated in this framework, lagging in accuracy compared to state-of-the-art (SOTA).

Efforts such as [8] that finetune a text LLM to improve an SAA output (processing is text only) underscore the value of leveraging textual contexts for speaker diarization, further motivating the use of speaker-aware LLMs to enhance SAA performance even further.

This work introduces a novel SAA framework that extends Granite-speech, a SOTA speech-aware LLM that held the top position on the Hugging Face ASR leaderboard at the time of its release [27]. We demonstrate its adaptation for SAA. Specifically, we show how a speech encoder optimized for ASR can facilitate accurate SAA with minimal modification and without degrading ASR accuracy (Section 2).

Our principal contribution, detailed in Section 3, involves enriching the speaker discrimination of the LLM by integrating both relative speaker tags and absolute speaker identifiers (actual identities or speaker classes). This technique yields significant improvements in speaker attribution accuracy.

In Section 4, we explain how augmenting training with artificially constructed conversations, in conjunction with real conversational data, boosts SAA accuracy even further.

In Section 5, we provide empirical results on both two-speaker and multi-speaker benchmarks, showing that our proposed methods surpass conventional diarization+ASR pipelines, and highlight the advantages of speech-aware LLMs for unified conversational understanding.

Finally, Section 6 presents concluding remarks and future directions.

---

* Core contributors

## 2. SPEAKER ATTRIBUTED ASR WITH SPEECH-AWARE LLMS

### 2.1. General framework

In speech-aware LLMs, speech input is encoded and projected into the text LLM's embedding space, thus allowing the LLM to process both speech and text within a single, unified architecture.

Our model is built upon Granite-speech-v3.3-8B [9], a speech-aware extension of the Granite-3.3-8b-instruct [10] language model. The speech input is first processed through an audio encoder, which is comprised of 16 Conformer layers. The encoder was originally pre-trained for automatic speech recognition (ASR) using a Connectionist Temporal Classification (CTC) objective. This encoder is followed by a down-sampling module and a Q-former projector (see Fig. 1).

The LLM is fine-tuned for a set of tasks including ASR, translation, and SAA, using a collection of training instances formatted as follows:

**Prompt**
<|audio|> *transcribe and denote who is speaking by adding tags such as* [Speaker 1]: *and* [Speaker 2]: *before speaker turns*

**Target Response**
[Speaker 1]: *And it's filling up with a*
[Speaker 2]: *yeah*
[Speaker 1]: *thing An alien presence*

where <|audio|> is a placeholder that is replaced with the projected encoded audio. During training, we keep the audio encoder frozen and optimize only the weights of the projector, while applying LoRA (Low-Rank Adaptation) [11] for efficient fine-tuning of the language model. This strategy preserves the original ASR capabilities of the encoder while allowing efficient adaptation of the LLM for SAA.

### 2.2. SAA using the CTC encoder

The CTC encoder was trained solely for ASR and thus encodes limited speaker-discriminative information in its output layer. As a result, its output provides insufficient cues for reliable speaker discrimination by the language model. While introducing a dedicated speaker encoder is a possible solution, it would introduce significant additional complexity. Alternatively, unfreezing and fine-tuning the original encoder degrades ASR performance.

Instead, we adopt a lightweight enhancement strategy: concatenating the encoder's output with that of an intermediate layer [12]. This combined representation is then passed to the projector. Our hypothesis is that intermediate layers, being closer to the raw audio, retain more spectral and paralinguistic information relevant to speaker discrimination. The selection of which intermediate layer to use is discussed in subsection 5.2.

## 3. EXPLICIT SPEAKER IDENTIFICATION FOR IMPROVED SAA

Training the LLM with relative speaker tags (such as [Speaker 1]) has a limitation: the model's learning is constrained to differentiating speakers only within individual conversations,

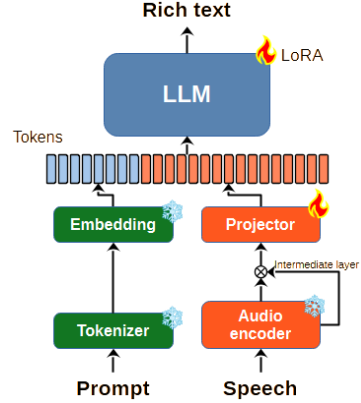

Fig. 1. Architecture of the Granite-speech model. Modules marked with a fire symbol are trainable during our fine-tuning process, while those marked with an ice symbol remain frozen.

preventing full exploitation of the training data. One approach, as described in Section 4, addresses this issue by augmenting the dataset with artificially constructed conversations assembled from existing speaker turns. In this section, we introduce a complementary and more direct solution: training the LLM to perform SAA in conjunction with speaker identification (SID). The corresponding target response is formatted as follows:

**Target Response**
[Speaker 1 ID 13259]: *And it's filling up with a*
[Speaker 2 ID 21794]: *yeah*
[Speaker 1 ID 13259]: *thing An alien presence*

where the ID number is a Personal Identification Number (PIN). However, speakers in test time are typically unseen during training, which may jeopardise the robustness of this approach. A possible remedy is to cluster the speakers in the training data into a set of speaker clusters, replacing each speaker's unique ID with a cluster ID.

Clustering is performed as a pre-processing offline step by extracting a speaker embedding for every speaker in the training and validation data, followed by clustering the speaker embeddings. We use the WavLM-ECAPA speaker embedding [13] available in the ESPnet toolkit [14] to extract embeddings for all the speaker turns in our training and validation data, and compute an average embedding per speaker.

The clustering step results in assigning a cluster index to every speaker in the training and validation data. We use $k$-means clustering [15] (with 100-300 clusters). Applying these cluster assignments to the target responses results in examples such as:

**Target Response**
[Speaker 1 cluster 14]: *And it's filling up with a*
[Speaker 2 cluster 52]: *yeah*
[Speaker 1 cluster 14]: *thing An alien presence*

## 4. DATASETS

The typical duration of speech input for speech-aware LLMs such as Granite-Speech is below 60 seconds. Therefore, training and evaluation were conducted on audio segments of up to 120 seconds in duration. We used both conversational datasets and synthetically created conversational datasets described in subsections 4.1 and 4.2 correspondingly. Table 1 lists the durations of the datasets we used.

### 4.1. Data preparation

**Fisher** [16] and **CallHome English (CH)** [17] are two-speaker telephone conversational dataset that are fully transcribed and include speaker labels. These datasets are widely used for training and evaluating SD systems. Each conversation ranges from 10 to 30 minutes in duration, and both datasets feature a significant amount of speaker overlap.

We created segments of approximately 10, 30, 60, and 120 seconds by selecting corresponding time spans from the audio and mixing the two speaker channels into a single mono track. Each resulting audio chunk is paired with a speaker-attributed transcript. Speaker overlap is handled at the annotation level using two simple heuristics:
1. Fully overlapping utterances (typically short vocalizations such as "uh-huh") were discarded from the transcript.
2. Partially overlapping utterances are handled by arranging the transcriptions sequentially.

**AMI-SDM** [18] is a multi-speaker meeting corpus frequently used for SD tasks. In our work, we use audio recorded from a single far-field microphone (Mic #1) and apply the same segmentation and preprocessing procedure described above. Similarly to Fisher and CH, we create 10-120s chunks.

**NaturalVoices** (NV) [19]: This dataset is based on podcasts from the MSP-Podcast Dataset [20] collection. The podcasts are segmented into speaker turns and transcribed using automatic tools. Using the provided metadata, we filter out segments containing either non-English speech or overlapping speakers. Then we extract continuous segments containing at least two distinct speakers, with durations up to 10-120 seconds.

**GALE** [21] is a dataset derived from broadcast news programs, with multiple speakers in each show. The GALE corpus is fully transcribed and includes speaker labels. We use a set of 20 English-language shows for evaluation. We chunk GALE into segments of 10–120 seconds containing up to four speakers.

### 4.2. Synthetic conversational data

To further increase the amount of training data, particularly for the multi-speaker scenario, we constructed the following synthetic SAA datasets:
- **Multilingual LibriSpeech** (MLS) [22] is a transcribed dataset of single-speaker recordings from thousands of labeled speakers. To create multi-speaker audio sessions, we randomly select 2–4 speakers. For each speaker, a random session is chosen, and short audio samples are extracted. These samples are then concatenated in an alternating pattern to simulate multi-speaker interactions, continuing until the desired segment length (10-120 seconds) is reached.
- **Fisher Augmented** (FisherA): We artificially construct conversations containing 3-4 speakers, by randomly selecting speaker-sides from Fisher and concatenating them as we did for MLS. We created 10, 30, 60 and 120s chunks. This data was used in addition to the original Fisher data.

*Table 1: Duration (in hours) of datasets we used.*

| Dataset | Train | Validation | Test |
|---|---|---|---|
| Fisher | 1838 | 60 | 60 |
| CallHome | - | - | 57 |
| AMI-SDM | 80 | 10 | 10 |
| NaturalVoices | 842 | 46 | - |
| GALE | - | - | 16 |
| MLS | 4774 | 143 | - |
| FisherA | 7821 | 300 | - |

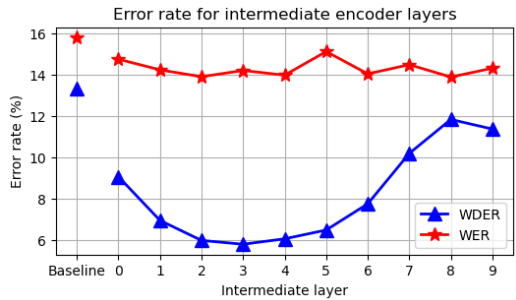

Fig. 2. WER and WDER as a function of the selected encoder layer. The baseline corresponds to using the encoder output without any intermediate layer augmentation. WER and WDER are averaged over various test datasets.

## 5. EXPERIMENTS

We report results for the Fisher, CallHome English, AMI-SDM and GALE test sets. We average results over test durations of 10, 30, 60 and 120 seconds. We use the training sets for projector and LLM finetuning, and use validation sets for model selection (stopping criterion).

### 5.1. Scoring

We evaluate SAA accuracy using the Word Diarization Error Rate (WDER), which measures the proportion of words incorrectly attributed to speakers, relative to the ground truth. We follow the protocol described in [3]. During evaluation, we consider only matched words, that is, words that are either correct or substituted, but exclude insertions and deletions from the calculation. Since speaker labels in SAA are relative (e.g., Speaker 1, Speaker 2), we compute the WDER after determining the optimal speaker label alignment between the prediction and the reference, to account for potential differences in speaker numbering.

### 5.2. Improving the CTC encoder

To identify the most speaker informative layer, we conducted a layer selection study, training multiple models using the output of a single intermediate encoder layer at a time. For each experiment, we extract the hidden representation from one of the encoder layers. This hidden representation is frame-wise concatenated with the encoder's final output and passed to the downstream projector module.

In unreported experiments, we found no benefit in combining multiple intermediate layers. The results of this layer selection process for the first ten hidden layers is shown in Fig. 2. Based on these results, layer 3 yields the best performance and is therefore used in all subsequent experiments.

*Table 2: Fine-Tuning the SAA model jointly with SID on the Fisher dataset. Reported results are in WDER (in %)*

| System | Fisher | CH | AMI | GALE |
|---|---|---|---|---|
| SAA | 1.5 | 3.7 | 12.8 | 18.5 |
| SAA + SID | 1.3 | 3.3 | 12.8 | 18.5 |
| SAA + SID (300 clusters) | **1.0** | 2.7 | 12.2 | 18.3 |
| SAA + SID (200 clusters) | **1.0** | 2.6 | **11.8** | 17.7 |
| SAA + SID (100 clusters) | **1.0** | **2.5** | 11.9 | 17.7 |

*Table 3: Fine-Tuning with different datasets. Reported results are in WDER (in %)*

| Training Data | Fisher | CH | AMI | GALE |
|---|---|---|---|---|
| SAA | | | | |
| Fisher | 1.5 | 3.7 | 12.8 | 18.5 |
| Fisher+MLS+NV+AMI | 1.2 | 3.0 | 12.1 | 12.4 |
| FisherA | 1.2 | 2.6 | 11.8 | 17.9 |
| FisherA+MLS+NV+AMI | 1.0 | 2.3 | 9.5 | **12.2** |
| SAA + SID (100 clusters) | | | | |
| FisherA+MLS+NV+AMI | **0.9** | **2.1** | **7.8** | **12.2** |

*Table 4: Comparison of our best SAA+SID method to SD based approaches.*

| System | Fisher | CH | AMI | GALE |
|---|---|---|---|---|
| WDER | | | | |
| PyAnnote+Whisper | 11.7 | 17.1 | 23.4 | 12.7 |
| PyAnnote+Granite | 11.0 | 15.1 | 19.7 | 12.7 |
| Nemo | 4.3 | 7.1 | 13.7 | **11.5** |
| Our best SAA | **0.9** | **2.1** | **7.8** | 12.2 |
| WER | | | | |
| PyAnnote+Whisper | 54.6 | 66.6 | 43.2 | 29.2 |
| PyAnnote+Granite | 41.2 | 42.5 | 39.9 | 30.6 |
| Nemo | 20.9 | 20.3 | 52.5 | **21.5** |
| Our best SAA | **17.7** | **17.8** | **22.9** | 22.6 |

### 5.3. Explicit SID for improved SAA

Table 2 reports the results of joint SAA and SID trained on Fisher. We observe that the joint SAA and SID system improves on two-speaker telephone conversations (which matches the training data) but not on the multi-speaker wideband test datasets. Clustering the speakers to 100-300 clusters has a positive effect and improves performance for all the test datasets.

For 100 clusters, WDER is reduced by 33%, 32%, 7% and 4% for Fisher, CallHome, AMI-SDM and GALE respectively. Regarding ASR accuracy, the average WER does not degrade (not in the Table).

### 5.4. Dataset ablation

We conducted a series of experiments using various combinations of training data. Selected results are reported in Table 2. The results indicate that adding synthetically concatenated conversational data consistently improves the performance over an already strong baseline. Training on both the conversational and artificially concatenated datasets reduces the WDER of the SAA system by 33%, 38%, 26% and 34% for Fisher, CallHome, AMI-SDM and GALE respectively.

Regarding ASR accuracy (not reported in the Table), WER decreased slightly on Fisher and CallHome, and decreased by 14% and 21% for AMI-SDM and GALE respectively. The AMI and GALE improvements highlight the importance of the inclusion of wideband training data. Training the joint SAA and SID model on the full training set reduces WDER even further.

### 5.5. Comparison to related works

To evaluate our SAA approach, we compare it against baselines that combine SD and ASR. Our first baseline utilizes the widely adopted pyannote.audio toolkit (speaker-diarization-3.1) [23, 24] performing SD, which segments the audio and assigns speaker labels to each segment. Each segment is then independently transcribed using an ASR system. The resulting transcriptions are concatenated, with speaker tags inserted to reflect the diarization output, producing a speaker-attributed transcript. For the ASR component, we evaluate both Whisper (large v3) [25] and Granite-speech (without fine-tuning).

For the second baseline we employ the NVIDIA NeMo [26]. NeMo runs SD and ASR independently, with the ASR component producing word-level timestamps that are then aligned to the segments from the SD system. We used the titanet_large model for diarization and the stt_en_conformer_ctc_large model for ASR.

To evaluate performance, we use the same scoring tools as in our main experiments, reporting both WDER and WER. Table 4. presents a comparison between the diarization-based baselines and our proposed SAA approach. As the results show, our SAA model consistently and significantly outperforms the diarization-based pipelines in both WDER and WER across all evaluated conditions (except for GALE, for which NeMo is better).

## 6. CONCLUSIONS

In this work, we introduced a novel framework for SAA by extending the capabilities of a speech-aware LLM. Our research demonstrates that our approach offers a more robust and effective solution compared to traditional pipelines that rely on separate SD and ASR systems.

We presented three key contributions to achieve superior SAA performance. First, we showed that the speech encoder, while initially optimized for ASR, can be effectively leveraged for speaker discrimination by concatenating its final output with an intermediate layer. This lightweight strategy significantly improves speaker attribution without degrading core ASR accuracy or requiring complex architectural changes. Second, we proposed a joint training methodology that integrates explicit speaker cluster identifiers with relative speaker tags. This technique proved highly effective. Finally, to overcome data scarcity, we demonstrated the benefit of augmenting existing corpora with artificially concatenated multi-speaker conversations.

Our empirical results show that our unified SAA model consistently outperforms conventional pipelines, including those based on widely adopted toolkits such as pyannote.audio and the NVIDIA NeMo framework. These findings underscore the potential of speech-aware LLMs for integrated conversational understanding.

Future work will focus on addressing current limitations, such as extending the framework to handle longer audio sequences. We also plan to investigate more sophisticated methods for speaker clustering.